# Engineering Gender-Inclusivity into Software: Tales from the Trenches

Claudia Hilderbrand, Christopher Perdriau, Lara Letaw, Jillian Emard, Zoe Steine-Hanson, Margaret Burnett, Anita Sarma[†]

Oregon State University
Corvallis, Oregon, USA
{minic, perdriac, letawl, emardj, steinehz, burnett, sarmaa}@oregonstate.edu

## ABSTRACT

Although the need for gender-inclusivity in *software itself* is gaining attention among both SE researchers and SE practitioners, and methods have been published to help, little has been reported on how to make such methods work in real-world settings. For example, how do busy software practitioners use such methods in low-cost ways? How do they endeavor to maximize benefits from using them? How do they avoid the controversies that can arise in talking about gender? To find out how teams were handling these and similar questions, we turned to 10 real-world software teams. We present these teams' experiences "in the trenches," in the form of 12 practices and 3 potential pitfalls, so as to provide their insights to other real-world software teams trying to engineer gender-inclusivity into their software products.

## 1 INTRODUCTION

Software has repeatedly failed diverse populations, falling short of aiding their productivity or even being usable by some populations [7, 8, 13, 22, 23, 26, 35, 43]. Such failures are serious: they marginalize people who "don't fit"—where "don't fit" can simply mean being different from the people who wrote the software. Of the many forms of diversity for which this problem arises, its connection with *gender* diversity is particularly well documented [4, 5, 6, 7, 8, 9, 10, 11, 13, 18, 22, 26, 27, 34, 35, 43, 44, 46].

Making software products equally usable to people regardless of their gender has practical importance—for both industry and open source software (OSS). If industry software teams fail to achieve inclusiveness, their market size shrinks. In OSS projects, if a project's tools or products fail to achieve inclusiveness, not only is product adoption reduced, but also the involvement of women and other underrepresented populations [17, 26]. Such loss of diversity matters to OSS teams, because with diversity comes better problem-solving, creativity, and excellence [20, 41].

A few methods have emerged to help software teams engineer gender-inclusivity into their software. One of these is the GenderMag method (Gender-Inclusiveness Magnifier) [10]. GenderMag is a method for finding—and, most recently, also fixing [43]—gender-inclusivity "bugs" in software. Empirical research reports that GenderMag is effective at helping software practitioners find and fix such inclusivity bugs [10, 43].

However, little is known about *how*—or even *if*—busy, real-world software teams can make such a method viable, given the many demands on their time and the practices they already have in place. To find out, we engaged with 10 software teams via Action Research, a type of longitudinal field study "that involves engaging with a community to address some problem... and through this problem solving to develop scholarly knowledge" [19].

Action Research is done collaboratively *with* participants—not "to" or "for" or "focused on" them. Therefore, our study was a fully collaborative endeavor with software teams who were working to engineer inclusivity into their software. As per Action Research's longitudinal focus, our involvement spanned months to years. Specifically, we had consistent involvement over 9 months with four professional software teams who create/maintain Oregon State Univiersity's Information Technology (IT), and intermittent data collection over periods ranging from 9 months to 3.5 years with six teams based in industry.

The results of this investigation contribute the first compendium of real-world software teams' practices and pitfalls in engineering inclusivity into their software, including:

- *Real-world practices* the software teams worked out for minimizing (time) *costs* of blending this method into their existing practices.

- *Real-world practices* the software teams worked out to maximize the *benefits and impact* they received for the time they spent using the method; but also...

- *Real-world pitfalls* the software teams ran into (and sometimes averted), potentially sabotaging their benefits.

- *Real-world practices* the software teams worked out to *leverage* and reap further benefits from the method.

- *Open issues* for which real-world practices are still emerging.

## 2 BACKGROUND

The practices we investigate are in the context of the GenderMag method, which empirical studies have reported to be effective [6, 10, 13, 34, 43]. We begin by summarizing GenderMag, a software inspection method for finding and fixing inclusivity "bugs".

GenderMag starts by helping a software team *find* user-facing inclusivity bugs in their own UI, using four "facets" of individuals' cognitive styles for going about problem solving. These facets form the core of the GenderMag method—an individual's *motivations, computer self-efficacy, attitude toward risk, information processing style(s),* and *learning style*(s).

GenderMag literature defines inclusivity bugs as issues tied to one or more of these cognitive facets. Such "bugs" are *cognitive inclusivity bugs,* but also *gender-inclusivity bugs* because the facets capture well-established (statistical) gender differences in how people problem-solve [2, 4, 5, 7, 11, 12, 14, 18, 22, 28, 35]. For example, using these facets, a software team might discover an inclusivity bug if a feature is easily discoverable by people with a tinkering learning style, but not easily discovered by people with a process-oriented learning style.

GenderMag makes the five facets concrete with a set of three faceted personas—"Abi", "Pat", and "Tim". Personas [1] are a widespread technique in industry. Each persona represents a subset of a system's target users—here, their purpose is to represent differences in the facet values. Abi's facet values represent the opposite end of the problem-solving style spectrum from Tim's, and Pat's facet values are a mixture of Abi's and Tim's. Without GenderMag usage, Tim's facet values are most often the ones software developers tend to design for, and Abi's facet values are often overlooked. Portions of the personas that are not about the facets (e.g., appearance, demographics, experience, job title, etc.) are customizable (Figure 1).

GenderMag sets these faceted personas into a systematic process via a specialized Cognitive Walkthrough (CW) [36, 45], as follows. Evaluators "walk through" each step of carrying out a use-case, and answer questions about *subgoals* and *actions* a user would need to accomplish those subgoals (italics added to show key differences from standard CWs):

SubgoalQ: Will *<Abi/Pat/Tim>* have formed this subgoal as a step to their overall goal? (Yes/no/maybe, why, *what facets are involved in your answer*).
ActionQ1: Will *<Abi/Pat/Tim>* know what to do at this step? (Yes/no/maybe, why, *what facets ...*).

ActionQ2: If *<Abi/Pat/Tim>* does the right thing, will s/he know s/he did the right thing and is making progress toward their goal? (Yes/no/maybe, why, *what facets....*).

As these questions show, identifying issues using this process includes identifying the *facets* that are tied with each. These facets are often key to the fixes—an issue's fix is designed around the facet that raised the issue. For example, to fix an issue that was raised for a particular problem-solving style, a team would revise that part of the UI to support *multiple* problem-solving styles: the already supported one and the unsupported one(s).

## 3 METHODOLOGY

To investigate the if's and how's of integrating GenderMag into real-world teams' practices, we worked with 10 professional software teams, 4 from the university and 6 from five companies. Our methodology for this investigation was Action Research.

### 3.1 The Action Research Methodology

Action Research [37] is a type of long-term field research, common in the fields of medicine and education and now emerging in various computing disciplines. Action Research has three stages: unfreezing, changing, and freezing [24]. In the *unfreezing stage*, an organization decides that a change is needed. In the *changing stage*, the organization experiments with new processes and creates variations with an eye toward producing the outcomes they want. The *refreezing stage* is when the new processes and changes become established as part of the organization's processes. The stages are not strictly linear; instead organizations often loop back to previous stages.

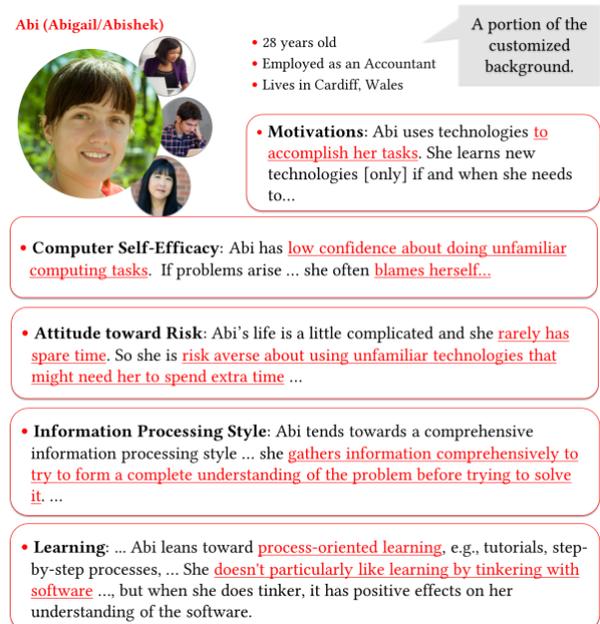

**Figure 1: Key portions of the Abi persona. See Fig 2 & 3 in the Supplemental Document for complete personas.**

Action Research is unlike many types of field research in two primary ways. First, it is iterative and "hands-on". Researchers work *together* with a community—the researchers are also participants, and the participants are also researchers [19, 37]. Second, its purpose is to develop scholarly knowledge about a problem to be solved, and also to iteratively solve it [24]. Thus, in contrast to other empirical methods, formative, summative, and treatment evaluations are intertwined within Action Research and cannot be separated.

Action Research emphasizes rigor by focusing on credibility and validity. Triangulation is widely used for this purpose; it reports phenomena only when multiple data sources, multiple data instances, and/or multiple investigators, etc., independently arrive at the same conclusions. Section 3.3 enumerates how our data collection processes facilitated triangulation, and Section 9 shows how triangulating these data cross-validated the practices and pitfalls we report.

### 3.2 Participants and Procedures

Our study included a diverse set of teams (Table 1). A mix of software developers, user-interface designers, site administrators, and marketing experts from Oregon State University and the five companies used the method on their own projects. About half the industry teams had previously used GenderMag, whereas most of the university teams were just starting. All the teams developed an interest in trying GenderMag (see Section 4 for more on this), and some contacted us about the method. A few used GenderMag on their own via the downloadable kit [9], but in most cases, they asked us to help them get started.

For teams who contacted us for help, we followed the same general process. Its main steps were: a pre-GenderMag meeting to show a team member how to customize a persona and help identify some suitable scenarios (use-case(s)) for analysis; and then a GenderMag session, which usually included time for debriefing. We started a team's first GenderMag session by briefly introducing the method's purpose, roles, and forms; and reminded them of the team's scenario (use-case) and customized persona. We then coached them through the session to whatever extent they wanted. After the first GenderMag session, we participated in later sessions only if a team asked us to; otherwise, teams continued (or not) on their own.

### 3.3 Data Collected and Analyzed

Central to our methodology's validity is triangulation, a cornerstone of qualitative analysis—whether the same results manifest themselves multiple times from multiple sources of evidence [33]. Toward this end, we collected data of multiple types to triangulate both within and among the teams. We also collected data from industry teams outside the university, to triangulate across multiple settings.

Table 2 summarizes the multiple data types collected from teams. From each team's GenderMag session(s), we collected the GenderMag forms they filled out, audio-recordings of the session(s) (which we then transcribed), the teams' customized personas, and our observers' notes. We also collected any artifacts we could, such as the teams' screenshots and/or mock ups. We then followed up with semi-structured interviews when possible, and in cases in which further data was offered (e.g., follow-up meetings, emails, public postings), we collected those too. For teams outside of our university community, we collected the same types of data to the extent permitted. When some types were not permitted or viable from a team's GenderMag session(s), we interviewed these teams. (The interview questions are listed in Fig. 9 in supplemental documents.)

We began our analysis of these data by listing all the potential practices that any team worked out, and any potential pitfalls they ran into, regardless of whether they found a way to avert it. As a validity measure, we then filtered out any practice/pitfall for which there was no triangulating evidence. Specifically, we required every practice/pitfall to have occurred in at least two independent occurrences or teams. Our purpose was to raise the likelihood that any practice/pitfall reported here would be potentially applicable to other real-world teams

**Table 1: The teams and number of team members from each who helped run GenderMag sessions.**

| Team name | Max # of members at session(s) | Applications these teams were working on |
|---|---|---|
| A | 6 | Information for instructors and students about <x> |
| B | Unknown | Interface for an AI product |
| C | 5 | Analytics and reports for staff to gain insights into university trends |
| L | 7 | <x> technologies |
| M | 2 | Education platform for instructors |
| N | >12 | An IT-support product for end users |
| O | 2 | Search engine |
| P | >7 | Web based interface for visual sorting with a deep learning back end |
| W | 3 | Web application for employees who manage <x> |
| Y | 7 | Application for customer communities |

**Table 2: We collected data from multiple sources for every team to enable triangulation.**
**Legend: Form=written forms filled out by the team during the session. Rec.=audio recording of session. Persona=the team's customized persona. Obs. notes=notes taken by observers.**

| Team | First GenderMag session | | | | More sessions | Other mtgs | Interviews | Emails, soc-media, shout-outs |
|---|---|---|---|---|---|---|---|---|
| | Form | Rec. | Persona | Obs. notes | | | | |
| A | ✓ | | ✓ | ✓ | ✓ | ✓ | ✓ | ✓ |
| B | | | | | | ✓ | ✓ | |
| C | ✓ | ✓ | ✓ | ✓ | | | | ✓ |
| L | ✓ | ✓ | ✓ | ✓ | ✓ | | | |
| M | | | | | | | ✓ | |
| N | | ✓ | | ✓ | ✓ | ✓ | ✓ | ✓ |
| O | | | | | | ✓ | ✓ | |
| P | | | ✓ | ✓ | ✓ | | ✓ | |
| W | ✓ | ✓ | ✓ | ✓ | ✓ | | ✓ | |
| Y | | ✓ | | ✓ | | ✓ | | |

looking for guidance on how to go about inclusivity-debugging their own real-world software.

## 4 FROM UNFREEZING TO CHANGING

From a diversity and inclusion perspective, Oregon State University had already reached the *unfreezing stage* (in Action Research). Oregon State had been placing increasingly greater emphasis on diversity and inclusion, and it had strong backing from the university's leadership. For example, Oregon State's latest strategic plan is structured under four primary goals, one of them being diversity, inclusion, and equity [32]. This strong interest by the leadership put them in a receptive state to think about changing their IT practices so that the *software* the university produces and uses would follow the same policies as the *institution*.

At this point, one of the authors of this paper approached the university's CIO with the idea of the IT organization making Oregon State's software gender inclusive. A few meetings with others in leadership positions ensued, and their awareness grew of the inclusivity issues that might lurk in their software:

*X-leadership: "Oh my God. What if the bias reporting software is biased?"*

The CIO's office decided to experiment with incorporating GenderMag into their processes. They funded one of the graduate students to help move it forward, began regular meetings, and arranged for the researchers to present the GenderMag method to a group of IT teams to see if any would want to step forward. We presented it at a campus IT meeting, and as Section 3 has mentioned, a number of teams expressed interest in trying it out. We report on those teams with whom we have the longest involvement.

The six industry teams in this paper were located in five companies at which the importance of diversity and inclusion had also been accepted. They had heard about GenderMag from presentations or papers and expressed interest in trying it.

These events brought the teams to the outset of Action

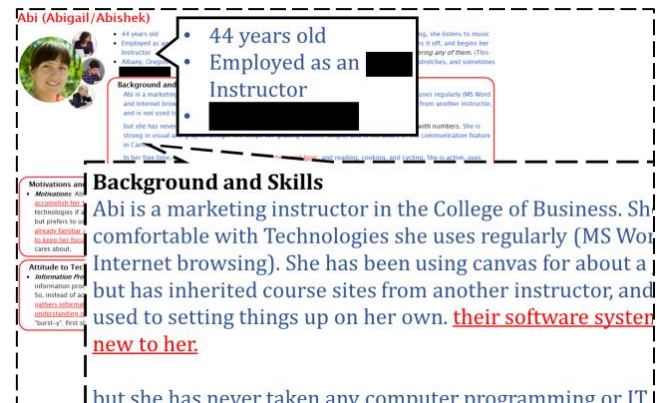

**Figure 2: Team A customized Abi to be an instructor by filling in the customizable parts of Figure 1. Blue text was customization, red text was fixed (not customizable).**

Research's *change stage.* For busy software teams, changes in process can be expensive, so teams needed to work out whether the upfront *costs* (time) of changing their processes to engineer inclusiveness into their software would pay off in useful and impactful *benefits*.

## 5 RESULTS: MINIMIZING SESSION COSTS

As Table 3 summarizes, the teams worked out three practices to help balance and/or minimize the cost of running their GenderMag sessions, which we detail next.

**Table 3: The teams' practices for minimizing their costs.**

| | Practices | Team |
|---|---|---|
| 1 | Designated Sub-team | A, M |
| 2 | Multi-path Evals | C, L |
| 3 | Evaluating UI Patterns | A, C |

## 5.1 Training vs. Efficiency and Follow-Through

Some teams ran into a trade-off between wanting to train team members vs. being efficient and maintaining follow-through. When first learning to use GenderMag, many teams included a large number of their team members in their initial GenderMag sessions. The potential advantages of such a large group can be that (1) more of the team gets (hands-on) experience with the method; and (2) more people in the room during the session potentially contributes more diverse perspectives during the evaluation, which can increase the completeness of the evaluation. These advantages come with a tradeoff of efficiency, since more team members would yield a longer discussion.

Team A was one of the teams who decided to include a large group in their first GenderMag session. They were evaluating a website for instructors and students (refer to Table 1). Team A focused on whether the information was easily findable by instructors and students with little spare time. They customized the Abi persona to be an instructor (Figure 2) and the scenario to evaluate: "Find instructions to add a TA to a course site."

For Team A, both of the above advantages materialized. Regarding the first advantage, Team A conducted their session with seven members. All seven members actively engaged in the session. The team's designated recorder took detailed notes, and some other team members taking their own notes as well. The second advantage materialized too. The relatively large size of the group helped bring out diverse perspectives, because the process captures the *union* of perspectives of *everyone* at the session—not just the more vocal people in the room. For example, Figure 3 shows one step of the evaluation in which some team members answered 'Yes' and others answered 'No'. With such a large group, the results were very thorough, ultimately identifying issues in 5 out of 14 (36%) of the evaluation steps they performed.

However, the large group size slowed down the evaluation: the more people's opinions to capture, the more time was spent on each question. During the entire session they finished only one scenario (14 evaluation steps), not the two scenarios the team had planned to evaluate. The team decided that this pace was probably too time-costly to be viable.

During a follow up meeting, the team decided to solve this problem by narrowing down the evaluation subteam to just three members. This also make clear who was accountable for following through on the issues they found (TA-2 refers to our transcription of Team A, line 2. TA-Email refers to an email message we received from Team A):

*TA-2: "...we are ... going to pair up based on whose people's time and availability align with moving forward"*
*TA-Email: "...we should be able to run through the full GenderMag process again with the two tasks above... it should provide a decent template for building a lot of the rest of the website."*

> **Practice 1:** *Designated Sub-team*
> Some teams narrowed an evaluation sub-team down to just a few team members, who kept the effort going through regular meetings and follow-through actions. This can reduce teams' time costs.

## 5.2 Walking Multiple Paths "At Once"

A GenderMag walkthrough, as a derivative of the CW [45], is designed to evaluate a single path (sequence of actions) through an interface—with no branching, because of the cognitive cost and group confusion of context switches between branches. However, Team C and Team L found that evaluating two small paths "at once" could increase their GenderMag method efficiency.

Figure 4 illustrates Team C's use of this practice. When

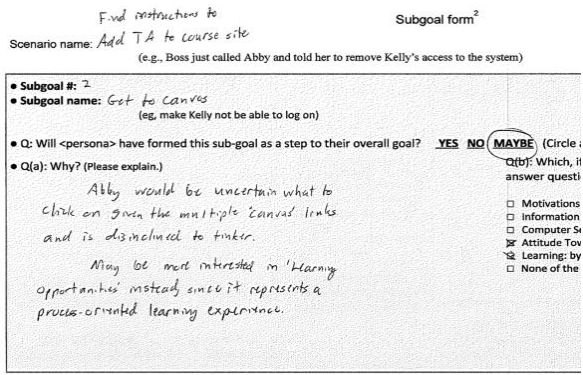

**Figure 3: Team A circled MAYBE to demonstrate that their team members did not all agree on either YES or NO.**

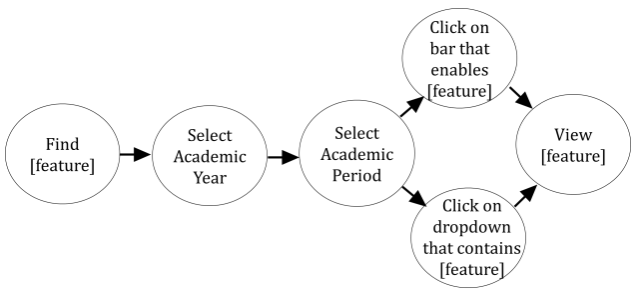

**Figure 4: Team C evaluated both of the small paths that Abi could take to reach the same subgoal.**

evaluating their software's analytical reporting "dashboards", they ran across two different paths a user might take from a single starting place to achieve a single goal. Both paths were short, and the team decided to evaluate both to compare them. Their multi-path evaluation paid off: they avoided re-evaluating in-common segments of the two paths. Their evaluation also revealed that the most straightforward path was not as discoverable as the alternative path, enabling the team to see why their users rarely chose the straightforward path:

*TC-364: "... there's two modes of getting to the answer here, so the first mode, she'd hover on the <feature>; it doesn't tell you what to do ... She's not going to realize she has to click on the bar."*

Similarly, Team L ran into multiple ways for their software to print a PDF. Comparing two possible paths with a multi-path evaluation like Team C's, Team L found an issue and a fix to make the most direct path discoverable to people with Abi's information processing style.

*TL-634: "...the image isn't linked, that would be nice... also there is more than one way to download the pdf; this is the most direct way..."*

> **Practice 2:** *Multi-path Evals*
> Teams that did "simultaneous" evaluations of two small paths could reduce the number of sessions needed to evaluate both paths. This practice was viable when the actions started and ended at the same place and achieved the same subgoal, and also facilitated direct comparison between the paths.

## 5.3 Reusing Evaluations

Teams A and C both worked out a way to generalize one GenderMag evaluation's results to other portions of their platform, making their GenderMag process more efficient and less expensive. They did so by evaluating a UI pattern that they were using in multiple places, and then applying their single evaluation's findings across all instantiations of that UI pattern in their software. For example, Team C selected a "representative" analytical reporting dashboard to evaluate, with the idea of applying their results across their application:

*TC-3: "...it's not just for one dashboard even though we tackled just one dashboard ... It's a good starting point for all our dashboards."*
*TC-6: "So some of the things we found in this session are definitely going to apply across the board..."*

> **Practice 3:** *Evaluating UI Patterns*
> Some teams selected a common UI pattern or set of related components for evaluation, and then reused their findings and fixes on other instantiations of that pattern, without having to run separate sessions for each.

# 6 RESULTS: MAXIMIZING SESSION BENEFITS

Teams worked out several ways to maximize the benefits they got from their GenderMag sessions, but also ran into potential pitfalls that could sabotage their efforts. Table 4 summarizes these practices and potential pitfalls, which we detail next.

## 6.1 Starting Early

Many modern software development processes recommend evaluating early in software lifecycles because of the reduced expense of fixing bugs early in the development process [3, 38]. Consistent with this recommendation, several teams used GenderMag early in their software development processes, using prototypes—sometimes paper-based or PowerPoint-based ones—instead of waiting for the software to be implemented. For example, Team Y saw the benefits of GenderMag'ing early:

*TY-117: "...<GenderMag> evaluating what we already have are ... excellent starting points ... we can begin to move the needle at really early points of design.... It's been really enlightening for me."*

> Practice 4: *GenderMag'ing Early*
> Several teams realized that using GenderMag early in the development process could ward off expensive changes to mature software and could also help them begin evaluating earlier in the software lifecycle.

## 6.2 Abi's Powers

The GenderMag kit [9] suggests that Abi provides the most powerful lens for finding inclusivity bugs; published GenderMag studies [8, 26] likewise report teams finding more inclusivity bugs with Abi. All of the teams followed this suggestion and used Abi as their first persona.

Team M, in using GenderMag on their web application for Computer Science instructors, had a second reason to use Abi first. Recall that the personas' demographics are customizable, so Abi can have any educational background, any profession, etc. Team M chose Abi to explore a tech-savvy user population who still had lower computer self-efficacy than their peer group:

*TM-14: "We chose to use Abi ... because we wanted to explore a user with low self-efficacy with the technology, ... it's hard to explain to our ... team members why somebody with multiple PhD's ... would blame themselves <for problems with the interfaces>"*

In contrast, Team N used Abi for the opposite reason: to find inclusivity bugs for users *not* trained in IT:

*TN-21: "we primarily relied on the Abi persona again, ... because we decided to err on the side of targeting... people who are expressly not IT people. <Abi's> attitude towards technology <risk> really tended to play a role."*

**Table 4: The teams' practices and potential pitfalls for maximizing their benefits.**

| | Practice or Potential Pitfall | Team |
|---|---|---|
| 4 | GenderMag'ing Early | O, W, Y |
| 5 | Abi First | A, C, L, M, O, W, Y |
| 6 | Abi = People! | C (and [8]) |
| | But Abi ≠ a Person | A, L, W |
| | Evaluating a Proxy UI | C, W |
| | Beyond our Control | C, L |

> Practice 5: *Abi First*
> All the teams used Abi as their first persona, perhaps because the literature reports Abi as offering the most powerful lens.

Despite Abi's powers, Abi is not all-powerful, and reflecting on the fact that Abi represents *people*—complete with human frailties—helped some teams gain insights. For example, Team C pointed out that, although some of their users train on the team's software, even trained users can forget what they learned:

*TC-398: "...people like Abi <are> who'll be using this, right? They ... spend half an hour train<ing>... do <this> for five minutes, then they go and do something else ... <They> can forget."*

Consistent with Team C, a previous GenderMag study [8] reported on two teams who ran GenderMag on the same software. In that study, the team who had gotten to know their users as people identified far more inclusivity bugs than the team who had never met their users. The latter team tended to assume that users would succeed at anything included in the users' training.

> Practice 6: *Abi = People!*
> Teams found reflecting upon the people their persona represents, who have human characteristics (including human frailties), enabled them to identify more inclusivity bugs.

## 6.3 Three Potential Pitfalls

Teams also stumbled across pitfalls that potentially threatened their sessions' likelihood of producing useful, actionable results.

The first potential pitfall was taking Practice 6 (*Abi = People!*) too far. Although the Abi persona represents the group of users with similar facets, some team members incorrectly personified Abi by attributing characteristics, beyond those in the persona document, of some person or people they knew. For example, in Team W, not everyone's understanding of Abi matched Abi's facet values. For example, one team member thought Abi would tinker, but tinkering is at odds with Abi's *learning style* facet—Abi learns by process, not by tinkering. The team member based their assumption on experience with real users:

*TW-329: "...they <Abi> are not going to pause, they are just gonna go and jab at it, that's what they do ... ten years of watching people do it tells me..."*

The team averted this potential pitfall because not all team members went along with the argument. However, had they all proceeded under the "Abi will tinker" assumption, they would be ignoring one of the facet values that has helped other teams find the inclusivity bugs they were looking for (e.g., [8, 26, 43]).

> Potential Pitfall 1: *But Abi ≠ a Person*
> Some teams noticed that assuming Abi is exactly like some real person a team member knows can backfire, resulting in evaluators taking into account fewer facets than they should be.

The second potential pitfall some teams encountered was running their evaluation on a "proxy" of the user interface, instead of the interface they were really interested in. For example, Team C wanted to evaluate an application that had recently been updated, but brought a machine to the session that didn't have the updated design. They tried to evaluate using this proxy, but problems arose: they had to pause and re-think because of features they saw that would not be in the new interface the users would see:

*TC-15: "...in the real environment, there wouldn't be all of these other tabs."*
*TC-20: "So it might not have the styling ..."*

Worse, the workflow and stylings that *were* available in the new interface to the users were *not* evaluated.

> Potential Pitfall 2: *Evaluating a Proxy UI*
> Teams who tried to evaluate a "similar" UI to the one they really cared about, ended up evaluating things that were present in the proxy, omitting things that were in the real UI but not the proxy, and/or spent extra time during the evaluation trying to keep the differences straight.

The third potential pitfall had to do with control and actionability: evaluating an interface the team has limited control over. For example, Team C encountered this potential pitfall with an unintuitive button for clearing a selection. They identified this as an inclusivity bug tied to three of the facets: *information processing style, computer self-efficacy, and learning style* facets, but then realized they couldn't fix it because it was a third-party element:

*TC-295: "...it's a <3$^{rd}$ party application> thing... we can't make it better. I wish we could. But we can't!"*

This potential pitfall can emerge in several situations: software that uses third-party APIs; software that is widely used, but not budgeted for redevelopment; and software that relies on sub-systems controlled by other teams [8]. Any of these situations can leave a team without the ability to act upon the results they find. In some situations, this is easy to avert (e.g., don't evaluate an interface unless the decision-maker(s) who "own" the system are present), in others, the system is so intertwined with other subsystems it can be difficult to avoid.

> Potential Pitfall 3: *Beyond our Control*
> The teams that tried to use GenderMag on interfaces or portions of interfaces that they could not change were less likely to gain any benefits from the evaluation.

## 7 RESULTS BEYOND THE SESSION

Teams also worked out practices that extended beyond the individual evaluation sessions, as Table 5 summarizes.

**Table 5: Teams' "beyond the session" practices.**

| | Practice | Team Name |
|---|---|---|
| 7 | GenderMag Moments | A, B, O, P, W |
| 8 | Debriefing | A, C, L, W, Y |
| 9 | Categorizing | A, C |
| 10 | Facet Survey | B, N, O, Y |
| 11 | Invite Abi | A, C, M, N, P, Y |
| 12 | Facets Drive Fixes | L, M, O, P, W |

### 7.1 GenderMag'ing in a Moment

Team N was first to tell us about a practice we'll term GenderMag Moments, but five teams ultimately used it. GenderMag Moments is a small fragment of a GenderMag session, triggered just-in-time by some kind of design question (e.g., "should we show the choices alphabetically or in the sequence they should be performed?") In a GenderMag Moment, team members already familiar with the full method, personas, and facets, answer the two GenderMag action questions in the context of the trigger:

ActionQ1: Will <*Abi/Pat/Tim*> know what to do at this step? (Yes/no/maybe, why, *what facets ...).*
ActionQ2: If <*Abi/Pat/Tim*> does the right thing, will s/he know s/he did the right thing and is making progress toward their goal? (Yes/no/maybe, why, *what facets....).*

For example, Team A started blending GenderMag Moments into their design meetings to consider how to fix issues they had found by using the full method. At first, they did not realize they were even doing so, until one team member pointed out:

*TA-31: "... we've just been doing Moments!"*

Team A also used GenderMag Moments in a slightly different way. They expanded them to include referring back to the GenderMag forms they had filled out originally, to make sure the design fix would address all inclusivity bugs that they had found.

> Practice 7: *GenderMag Moments*
> Teams worked out two versions of GenderMag Moments: (1) using the GenderMag questions to guide the evaluation of design solutions just-in-time; (2) Using the earlier sessions' filled-out forms to evaluate whether the fixes would address all the inclusivity bugs they had originally identified.

### 7.2 Reflecting Back and Getting Organized

Reflecting back at the end of a session was a common practice, with five teams using it to good effect. For example, in reflecting back upon their first session, Team C realized they could address some inclusivity bugs they previously thought were unfixable due to third-party software limitations. Ultimately, they found a way forward using the third-party software in ways they had not thought of before. In fact, Team C found their debrief so valuable, they scheduled a follow-up meeting to continue:

*TC-684: "So I think what we have to schedule another meeting, right, kind of follow up meeting after people have had a chance to think about what we saw here today..."*

A particularly useful way two teams spent their debriefings and follow-up sessions was organizing the discussion outcomes and inclusivity bugs they had found into categories. Team A categorized inclusivity bugs by navigation level: the homepage ("first") layer, and the next click in ("second layer") (Figure 5).

*TA-6:"...the next layer for today that... we wanted to tackle was, assuming that the homepage looks okay... how do we lay out information in a second layer..."*
*TA-6: "So that is what <team member> has been mocking up is once we get past the first layer..."*

In contrast, Team C categorized inclusivity bugs by the type of remedy they felt would address the bug.

*TC-17: "... talk about which bin it would go under... a training thing... a styling consistency thing or an in-figure key."*

For both teams, categorizing in this way helped the team decide how to and who would address which inclusivity bugs.

> **Practice 8:** *Debriefing*
> Many of the teams debriefed after the GenderMag session to discuss actionable tasks, next steps, insights, and workload.

> **Practice 9:** *Categorize Issues*
> Splitting inclusivity bugs into categories helped some teams develop action plans for fixing them, evaluate feasibility of the fixes, and/or gauge the amount of effort needed to implement fixes.

### 7.3 Surveying Real Users' Facets

Like other inspection processes, GenderMag sessions are a *complement* to empirical studies with users—just as inspection processes like code inspection are a complement to testing with real data. In doing such user studies, four teams worked out multiple ways to leverage GenderMag via survey questions to gauge the facet values of their users and/or participants.

This practice started when Team N decided to do a survey to find out what facet values their own user populations had. Team N had a history of using surveys to categorize their user populations, so they merged portions of their existing surveys with questions like the one in Figure 6. Some of the questions they added (including the ones in Figure 6) came from literature searches for validated questionnaires, and others had to be worked out from scratch.

Team N later used the same survey to recruit participants for in-person field studies. Recall from Section 2 that people use diverse problem-solving styles. Team N's goal was to cover a span of this diversity in an upcoming study, so they administered the survey during recruiting and selected participants that spanned its range of results. Team N later shared their facet questions, and Team O also started using them

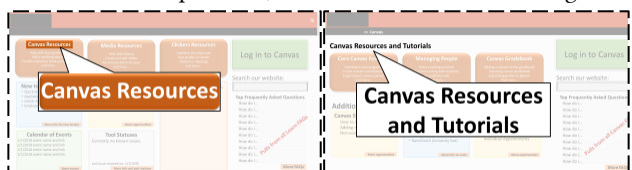

**Figure 5: (Left)** First level of Team A's software fix. **(Right)** Second level, after clicking on "Canvas Resources" link on first level.

to recruit for a lab study:

*TO-Meeting: "...we can have the '<potential> participants' fill out the short survey question – then we can find out... the facets that we miss"*

Team B and Team O then started using the facet questions to help *analyze* data from their lab studies. For example, Team O grouped the inclusivity bugs they found by the facet values that had revealed them. This helped guide their work toward fixing these inclusivity bugs (for all personas, not just Abi)—and to then measure whether the fixes actually made their system more inclusive. Their lab study revealed that the resulting system was indeed more inclusive and was generally as good or better than the original across almost *all* of the facet values.

> **Practice 10:** *Facet Survey*
> Teams used survey questions measuring people's facet values in multiple ways: (1) to understand their user populations, (2) to recruit for user studies, (3) to analyze their lab study data, and (4) to measure the effectiveness of fixes.

### 7.4 Beyond the Session with Abi & the Facets

As per recommendations from persona research [1], six teams found ways to bring Abi (and sometimes other personas) into the workplace. The goals behind this practice were to remind themselves to keep Abi in mind, and to help their coworkers ground their conversations in Abi's attributes. Abi turned up on desks, in presentations, on posters in the lab, and even on "Hello, my name is..." nametags for meeting attendees:

*TA-29: "So I did have <Abi persona> paper on my desk, right next to me."*
*TC-203: "We'll have the Abi persona with us when we're doing presentations."*
*TN-4: "Whenever I do GenderMag I make people wear little tags that say: 'I am Abi'... So, everybody remembers they are not themselves"*

In the contexts of their evaluation sessions and design discussions, several of the teams also learned to regularly refer to Abi by name and to refer implicitly to Abi's facets:

*TA-107: "I guess the principles of navigation for Abi are, as long as... <Abi is> confident <Abi is> moving through <the> path it's not bad to necessarily have an additional click."*

Five of the teams took Abi's facets one step further: they used the *facets* to engineer the *fixes* to the inclusivity bugs they found. For example, Team O fixed the UI widget in Figure 7 to better support Abi's motivations and risk facets (recall Figure 1). They removed the counts (the right side of each bar) to make the filters look more like filters, so that if a task-motivated user like Abi was trying to filter, they would see that widget as the way to accomplish their task.

**Figure 6. A portion of the facets survey used by some of the teams. This portion measures computer self-efficacy. The complete survey can be found in Fig. 1 in the supplemental document.**

---

**Practice 11:** *Invite Abi to the Office*
Most teams found ways to keep Abi (and the other personas they used) in front of themselves and their coworkers. Among their practices for doing so were pictures on their desks, posters, nametags, pictures in their slide presentations, and in regular conversations.

---

**Practice 12:** *Facets Drive Fixes*
Some teams used the GenderMag facets as ways to work out their fixes, and as reasons to explain specific changes to their colleagues, which helped spread awareness about how cognitive styles were being left out and how to fix UIs to correct that.

---

## 8 DISCUSSION: HEATED DISCUSSIONS IN THE TRENCHES

The practices that we observed have helped GenderMag gain traction at Oregon State University and at other organizations. Still, these practices do not address everything that can arise in the trenches. Here we discuss two issues that some teams faced and emerging ways to potentially address them.

### 8.1 Arguing over the Use-Case Sequence

Earlier versions of the GenderMag method required a team member to pick out an exact sequence of actions to be evaluated in advance, as with the traditional CW [45] (a parent of the GenderMag method). This did not lead to arguments, but the pre-work required was shown in field studies to be burdensome and mostly unnecessary [6, 8].

Thus, by the time of the current investigation, the GenderMag process had evolved so that the only pre-work required was to customize the persona (if desired) and name the use-case(s) to be evaluated. The specific action path through the use-case was left to the team to choose just-in-time, one action at a time, as the session progressed.

This led to a new problem. Some team members had very different ideas about which action path to evaluate, debating at length during the session the next step to evaluate. Such debates have arisen for multiple teams, consuming time and leading teams to even try to backtrack-modify entire use-cases midstream, leading to ever more confusion.

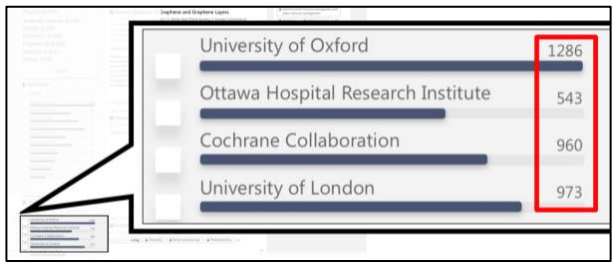

**Figure 7. The filtering widget originally included a counts column (right). The team decided that task-motivated users like Abi might not see it as a filtering device, since it looks more like statistics. Team O removed it.**

To avoid this problem, we started coaching teams to leave to the prototype's UI "driver"—the person who does the actual clicking through the prototype during a session—the decision of what next step of the sequence to evaluate. So far, arguments over the next step in a sequence have not been reported or observed since we made this change.

### 8.2 Sometimes Talking about Gender is Hard

Gender biases and their implications can be a controversial topic, and some team members eager to *make* their software less biased were still not comfortable *talking* about gender. To those team members, the name "GenderMag" was uncomfortable:

*TM-10: "I think the name GenderMag was kind of distracting. I had to clarify to people that it's about gender differences but that's not the only important part of it."*
*TB-64: "... I would be happier with a different name. But I didn't come up with one."*

This discomfort echoes earlier reports of teams wanting to "talk about gender without talking about gender" [6, 30]. Some have resolved it by adopting the vocabulary of the facets instead (e.g., different levels of risk tolerance, information processing styles etc.) [6]. Another solution arose during the time of this investigation—referring to GenderMag's "family name" instead, InclusiveMag. Early feedback on this alternative has been encouraging, but we have not yet seen it in the field.

## 9 THREATS TO VALIDITY AND MITIGATIONS

No empirical study is perfect. One reason is the inherent trade-off among different types of validity [48].

*External validity* refers to the ability to generalize the findings of a study. We mitigated the risk of introducing threats to external validity by analyzing multiple teams in a university and in industry. Even so, the practices that we collected from the teams may limit our ability to generalize the use of these practices to teams outside these groups.

*Internal validity* refers to how the study design can influence conclusions of the study. Our study has several uncontrolled variables. For example, as an Action Research study, we did not attempt to control for teams' prior design practices or knowledge of gender issues; even had we wanted to, there is a lack of robust measurements for either. Teams and team members varied in the levels of insights they were able to gain from the method; some of these variations could have been due to the members' pre-existing ability to empathize with their users, and some could have been due to the project each was evaluating. There were also several factors that may have determined what we did and did not observe, such as team members' prior experience with inspection methods and the make-up of the teams and projects. Therefore, some of the interpretations we made from the data might be different had we studied different teams or projects. Finally, as in any Action Research study, we worked with the teams to help them develop solutions. This impacts the replicability of our results.

Field studies, including Action Research studies, achieve real-world applicability, whereas controlled studies achieve isolation of variables. To reduce effects of the threats above, we collected data from multiple teams and software projects and made extensive use of data triangulation, as detailed in Table 6.

## 10 RELATED WORK

Although research into accessibility (e.g., [47])—which aims to improve ability-based inclusivity of software, such as accessibility for low-vision people—is long-standing, most other forms of software inclusivity have started receiving attention only recently. Still, in the last decade, the importance of inclusiveness and diversity in software has sparked interest in the research community and industry. This has led to new conferences and conversations to address biases in software [15, 21, 27, 31, 42, 46].

Online communities can exist only via software, and several research groups have investigated gender diversity and its effects on online communities [29, 39, 40, 41, 44]. For example, Vasilescu et al. found that diversity within OSS communities, while limited, helped strengthen codebases [41]. Ford et al. found that "peer parity" (having similar others for comparison) was an important factor in women's decision to engage in a software development community [16]. Mendez et al. found that gender biases in OSS tools and infrastructure can impact OSS newcomer success [26]. Terrell et al. found that, among new contributors (non-core members/outsiders), men's and women's pull request acceptance rate was similar when their profiles are gender-neutral but gender-biased when gender could be identified [39]. Such inclusivity bugs are problematic for both an organization's community and its productivity, as research across multiple fields has repeatedly shown. As a recent example in software engineering, Vasilescu et al.'s analysis of GitHub software projects and participant surveys found that gender and tenure diversity significantly increased productivity [41].

Outside of gender-inclusivity, other research has investigated other inspection methods in real-world settings, such as heuristic evaluation and CWs; one notable example is [25]. However, these methods, and therefore investigations of their use, are not about engineering *inclusivity* into software.

As far as methods for identifying, preventing, and/or fixing inclusivity bugs in software, there is only a little research. One such work is the GenderMag method, summarized in Section 2. The other one we know of is Williams' collection of design process recommendations for including women in the decision-making that shapes software [46]. However, there has been almost no investigation of how to integrate such methods into a real-world setting that already has longstanding software engineering practices in place. The above two papers and [6] are the only works we could locate on this subject. This paper helps to fill this gap.

## 11 CONCLUSION

In this paper, we have presented a longitudinal field study in which ten real-world software teams at six different institutions worked to "engineer inclusivity" into their software. The investigation spanned from 9 months to as long as 3.5 years in one team's case. The results revealed 12 practices, 3 potential pitfalls, and 2 issues the teams worked on or encountered in combining the new method with their existing team practices and cultures. Some of the particularly novel practices they worked out were:

- Even though GenderMag is an *inspection method*, teams used it to re-invent their ways of recruiting for and analyzing some of their *user study methods*—by leveraging the method's facets into survey and analysis instruments (Practice 10).

- Even though GenderMag operates at the level of concrete UIs, teams found a way to *abstract above them* to *UI patterns* that were common in their applications (Practice 3).

- Even though GenderMag is a systematic process that traverses *an entire use-case*, teams invented a way to do *just a "moment"* of the GenderMag process, just in time to make design decisions while working out their fixes (Practice 7).

**Table 6: Evidence behind each practice/pitfall. The checkmarks are instances of the data sources (columns) providing the evidence. For example, the Debriefing practice was in 5 initial GenderMag sessions, 1 multiple-GenderMag session sequence, and 1 follow-up meeting.**

| | First GM session | Multi GM sessions | Follow-up mtgs | Interviews | Emails | Evidence in prior lit. |
|---|---|---|---|---|---|---|
| *Minimizing Costs* | | | | | | |
| 1 Desig. sub-team | ✓ | | | ✓ | | [6] |
| 2 Multi-path evals | ✓ | | ✓ | | | |
| 3 Eval UI patterns | | | ✓✓ | | | |
| *Maximizing Benefits* | | | | | | |
| 4 GM'ing Early | ✓✓✓ | | | | | [6,43] |
| 5 Abi First | ✓✓✓✓✓✓✓ | | | | | [8,26] |
| 6 Abi = people | ✓ | | ✓ | | | [8] |
| Eval'ing proxy | ✓✓ | | | | | |
| But Abi ≠ person | ✓ | ✓ | ✓ | | | |
| Beyond control | ✓ | ✓ | | | | [8] |
| *Beyond the Session* | | | | | | |
| 7 GM Moments | | | ✓ | ✓✓✓✓ | ✓ | |
| 8 Debriefing | ✓✓✓✓✓ | ✓ | | | | |
| 9 Categorizing | | | ✓✓ | | | |
| 10 Facet Survey | | | ✓ | ✓ | ✓✓✓ | [43] |
| 11 Invite Abi | ✓ | ✓ | ✓✓✓ | ✓✓ | | [6] |
| 12 Facets Drive | ✓ | | | ✓✓✓✓ | | [43] |

# Engineering Gender-Inclusivity into Software

This paper is the first extensive investigation into practices of real-world teams who were exploring how to go beyond just making their software work, to making it work equally well for different genders. Perhaps the central message behind these teams' experiences is that suspecting your software of gender-bias and wanting to fix it are all very well and good—but integrating a systematic process can make all the difference:

*TC-3: "I thought it was very, very informative ... there are some things that we <already> knew we had to change ... but this ... gave us a process"*

## ACKNOWLEDGMENTS

Oregon State University and industry collaborators.